# What Lies Between Crystal and Randomly Packed Structures? A General Characterization of Non-Periodic Order

Ian M. Douglass and Peter Harrowell*

*School of Chemistry, University of Sydney, Sydney New South Wales 2006 Australia*

* Corresponding author: peter.harrowell@sydney.edu.au

Abstract

In this paper we address the characterization of the structure of condensed materials, periodic and non-periodic. Carrying out an extensive study of over 7000 different groundstate structures of a 2D lattice model of binary packing, we find a predominance of non-periodic structures (over 96%) that extend across the entire range of possible diversities. These non-periodic structures are resolved by establishing whether a structure will accommodate or reject additional local structures. This property, structural selectivity, is treated as a signature of an underlying ordering principle. The major result of the paper is the determination that roughly 35% of the non-periodic structures are selective and, hence, ordered in some way. This selectivity extends up to a diversity of ~ 9, well beyond the upper threshold for diversity in periodically ordered states.

## 1. Introduction

Crystals (at least many of them) are characterized by structures of low complexity while amorphous solids or glasses are generally regarded as structures of maximal complexity. Aside from the obvious issue of how are we defining complexity, this observation invites the question, what types of structures occupy the range of complexities that lie between these two limits? How many such structures might exist and how can we characterize them? Have we been inadvertently bundling quite different classes of structures together under the omnibus heading ‘amorphous’ for lack of the means of differentiating them? These sorts of questions motivate the work presented in this paper.

The idea that the complexity of a structure is a property of physical significance dates back, at least, to Pauling’s proposition in 1929 that the structure of ionic crystals reflected a tendency to minimize the number of different coordination geometries (the so-called rule of parsimony) [1]. The number Z of atoms in the unit cell provides an obvious starting point in quantifying the structural complexity of a crystal. This number, however, retains redundant information since the unit cell will typically contain a number of sets of sites, with each set consisting of sites related by a symmetry operation of the crystal space group. Each set of sites is identified by a representative site referred to as a Wyckoff site. We can, therefore, reduce the unit cell to the asymmetric unit, a subvolume of the cell that contains only one site from each of the symmetry related sets. As suggested by Baur et al [2] in 1983, the number of atoms in the asymmetric unit provides an improvement over Z as a measure of complexity. In 2012 Krivovichev [3] defined a measure of crystal complexity using a Shannon entropy based on the number of symmetry-defined sets of sites (the crystallographic orbits [4]) and the number of sites within each orbit, an expression we shall return to below. The research into crystal complexity has been extensively reviewed by Krivovichev [5].

Since all the measures of crystal complexity make explicit use of the size and symmetries of the unit cell, they cannot be directly applied to non-crystalline materials. A more general approach to structural complexity was proposed by Kurchan and Levine (KL) [6] in 2011. KL reasoned that in an ordered system there must be patches of particles whose structure repeats more frequently than expected from a purely random assembly. The complexity of a structure could be measured by determining the frequency with which patches repeated. In 2019, Martiniani, Chaikin and Levine [7] introduced an explicit measure of this complexity. Starting from the observation that "the more ordered a system is, the shorter the description required to specify a typical microstate", they defined the computable information density (CID),

$$CID = \frac{L_c}{L_o} \qquad (1)$$

where $L_c$ is the length of the compressed data string while $L_o$ is the length of the original uncompressed string (the latter being proportional to N, the number of particles in the structural sample). For any structure characterised by long range order, $L_c$ exhibits a less than linear dependence on N so that, in the limit of large N, the CID vanishes, irrespective of whether that long range order is periodic or aperiodic. This feature of the CID, along with the unavoidable disruption of local structural correlations arising from the mapping to a 1D string as part of the compression algorithm, means that this approach to complexity, while an effective tool for establishing extensive structural correlations and the associated decrease in configurational entropy [8,9], is not well suited to resolving the different patterns of local structural arrangements that distinguish many-body groundstates. To address the question of structural differentiation, we introduced, in 2019, the concept of structural diversity [10]. Borrowing from the ecological measures of species diversity [11], structural diversity is defined as an effective number of local structural 'species' that constitute an extended configuration. Local structures have long provided the basis of structural characterization in liquids and amorphous solids. There exist a number of excellent reviews on this subject [12,13]. The general strategy in this approach has been to identify a small number of favored local structures, selected for their frequency [14], temperature dependence [15], or match with some target geometry [16], and then to use the concentrations of these selected local structures to account for the observed properties of the disordered material [12,13]. In introducing the structural diversity, we are shifting this focus from the topological/geometrical details of a few selected local structures to look, instead, at the effective number of local structures contributing to a given configuration. As we shall discuss in Section 3, structural diversity can be regarded as a natural generalization of Krivovichev's measure of crystal complexity.

As our goal is to develop a general characterization of order, it is important that we understand the relation between complexity and order. KL [6] explicitly equate the two concepts: the increase in the compressibility of structural data is directly linked to the spatial extension of a structural correlation length. The emphasis by KL on the length scale of correlations, as opposed to the description of the correlations themselves, is an important feature of their approach. The diverging length scale is an *attribute* of order. It is an attribute of particular interest to KL as it is directly linked to the persistence of structure and, hence, solidity. Diversity, in contrast, provides a well resolved descriptive characterization of structures but one that is only loosely correlated to order (as we shall demonstrate). In general, an ordered state is one that obeys some sort of underlying ordering principle [17]. Among the large number of possibilities, these principles might include broken symmetries and/or the periodic repetition of structural motifs, features that are easily detected. It is not true, however, that one can always determine the ordering principle through simple observation of the structure. Uncertainty remains, for example, regarding the ordering principle that underlies quasicrystals [18] with a number of proposals – tilings with matching rules [19], maximal entropy random tilings [20] and cluster density maximization [21]– under consideration. Here, then, is where the attributes of order enter the picture. Presented with a structure that lacks a broken symmetry or a regular repetition, we are left with the identification of an attribute of order as our signature of the existence of an underlying ordering principle.

So, what are some typical attributes of order? *Structural persistence*, if defined as the divergence of the persistence time at a non-zero temperature, implies, as noted by KL [6], the divergence of a correlation length of some underlying order. A second attribute of order is *structural selectivity.* A specific ordering principle must, by definition, exclude some structural options. We can, therefore, detect the presence of such a principle by demonstrating that potentially stable structural options are excluded from the groundstate. In Section 4 we shall develop an explicit protocol for determining whether a given groundstate structure is selective or not. While we might reasonably expect low diversity structures to be selective and high diversity structures to be non-selective, we know of no explicit connection between the two properties and so, in this paper, we shall treat complexity and order as two distinct, although possibly correlated, features of a structure. In the following discussion we shall only refer to structures as being ordered when we can explicitly identify an ordering principle as in the case of periodic groundstates. If, instead, we rely on an attribute such as selectivity, we shall classify structures as selective or non-selective.

This paper is organized as follows. To explore the utility of different measures of complexity and order, we need a large space of structures that span the range from simple crystals to random arrangements of particles. To this end, we introduce, in the following Section, a lattice model, the Favored Local Structure (FLS) model, that can efficiently generate a large and diverse range of groundstates. In section 3 we shall resolve the distributions of the periodic and aperiodic groundstate structures of the FLS model using structural diversity. In Section 4 we introduce an explicit prescription for identifying structural selectivity and apply it to our distribution of groundstates to establish just how far ordering extends beyond the periodic subset of structures.

## 2. Model and Methodology

We need a model material which allows for a direct adjustment of the structural complexity and whose groundstates can be efficiently calculated. In this paper we shall use the Favored Local Structure (FLS) model, introduced by Ronceray and Harrowell [22] in 2011. The model consists of a two-dimensional triangular lattice with each site occupied by either an A particle or a B particle. A 3D version has also been studied [23]. The local environment of a given site is defined as the pattern of A or B occupancy of the six nearest neighbor sites. There are 14 distinct local structures, i.e. LS's that cannot be interconverted by a rotation as shown in Fig.1. We can then define a specific *system* by selecting any one or more of these distinct local structures to be stable and assign them an energy of -1 (the 'favored' local structures) with the remaining structures assigned an energy of zero. This assignment of energy to a configuration constitutes the many-body Hamiltonian of the model.

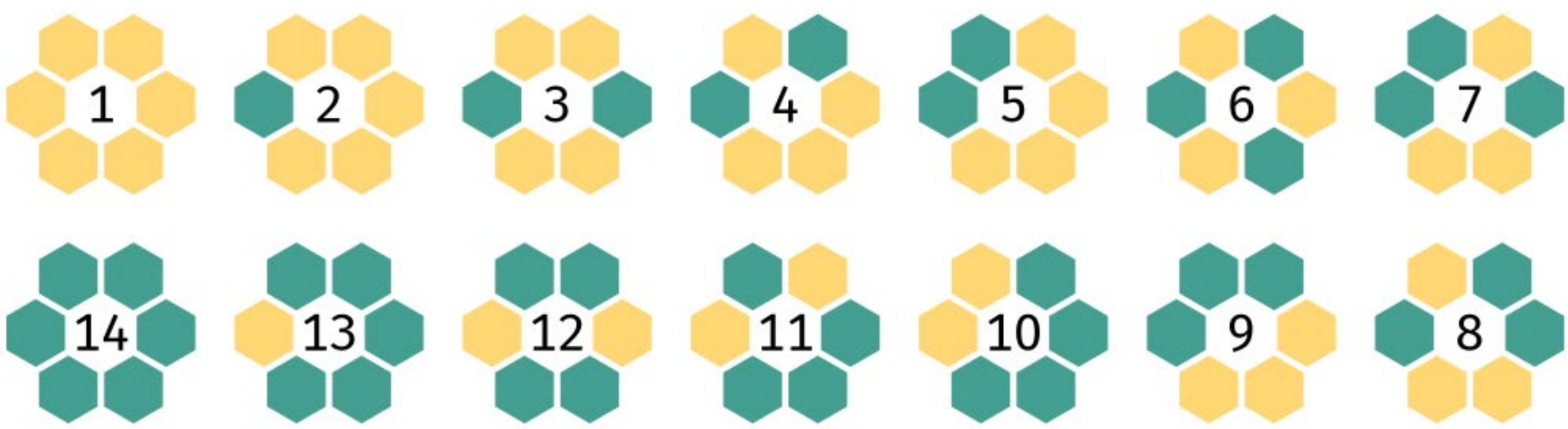

**Figure 1.** The 14 distinct local structures (LS) in the Favored Local Structure model on the 2D triangular lattice. The local structures are referred to in this paper using the labels as shown in the figure.

There are $2^{14}$ possible systems, each characterized by a particular choice of the set of favored local structures (FLS). We have generated groundstates for these systems by cooling NxN simulation cells, for $16 \leq N \leq 64$ and N =256 from T = 1.6 to T = 0 at a rate of 5 x $10^{-6}$ per Monte Carlo (MC) cycle. The restriction to groundstate structures avoids the complication of structural analysis in the presence of thermal fluctuations. For each system, the smallest crystal (if one existed) was taken as the 'canonical' ground state, otherwise the 256x256 state was used. The result of this quenching protocol is a total of 7609 unique groundstate structures. Each groundstate can be characterised by its composition defined by those local structures that occur with a frequency $p > 0.015p_{max}$ , where $p_{max}$ is the value of the frequency of the most abundant local structure found in that groundstate. (The choice of the factor 0.015 is made such that we would successfully identify all of the contributing local structures in all of the crystal phases generated in the lattice model. Note the cutoff in p is not used in the calculation of the diversity $S_1$.) In this paper we shall characterise a groundstate by the list of constituent local structures defined in this way. Included in these groundstates, we found 287 periodic structures (note that we have included as distinct crystal structures that differ only by a swapping of all particle identities). Periodic structures were identified by testing for the existence of a repeated unit cell with sides parallel to those of the simulation box. The unit cell used to calculate Z for each periodic structure is the smallest possible repeat unit with no constraints on the orientation of the basis vectors. We find the periodic structures cover a wide range of Z, the number of particles in a unit cell, as shown in Fig. 2. Examples of the minimal unit cell crystal structures, taken from across the range of Z values, are shown in Fig. 3.

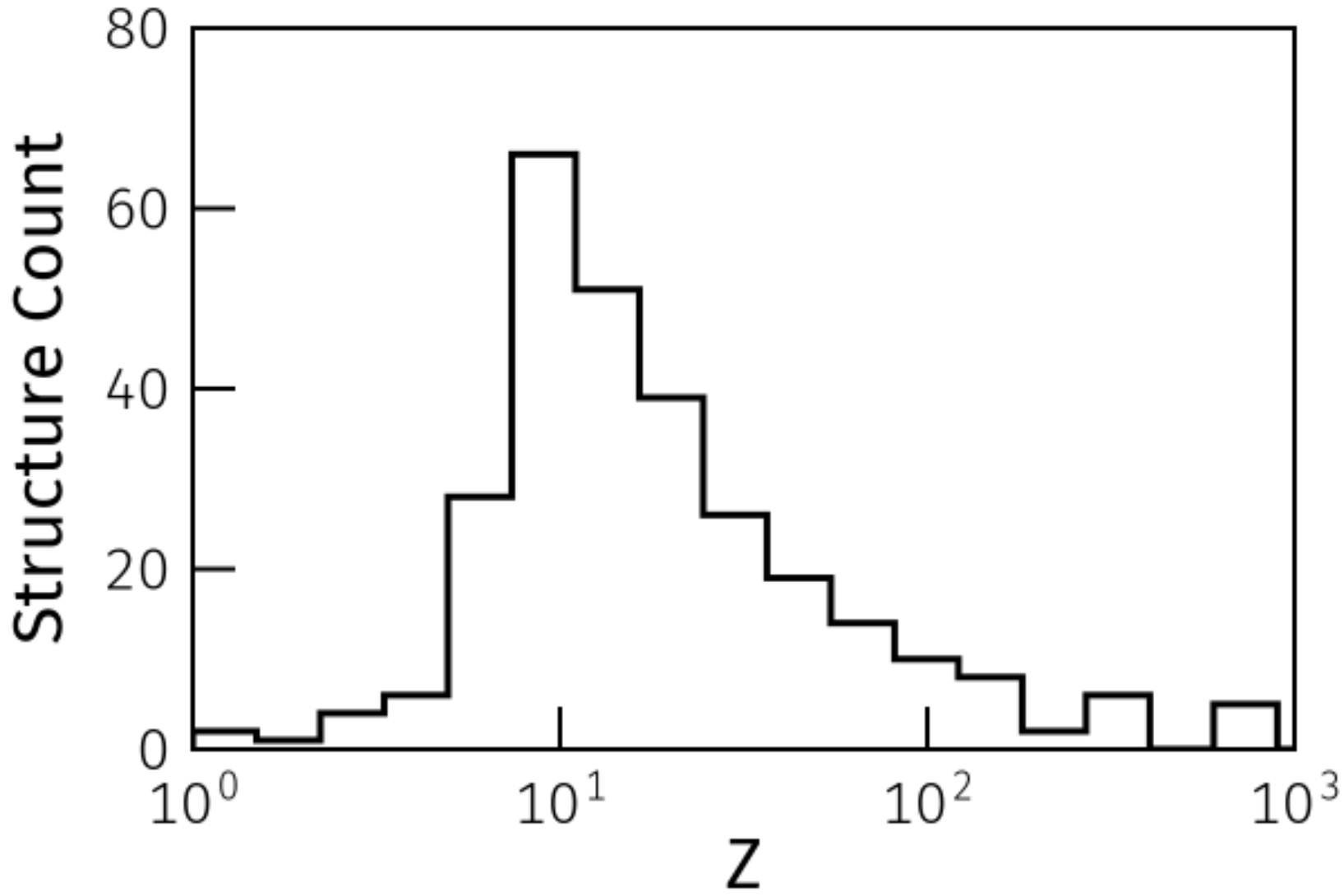

**Figure 2.** The distribution of the unit cell size Z for the 287 crystal groundstates of the FLS model.

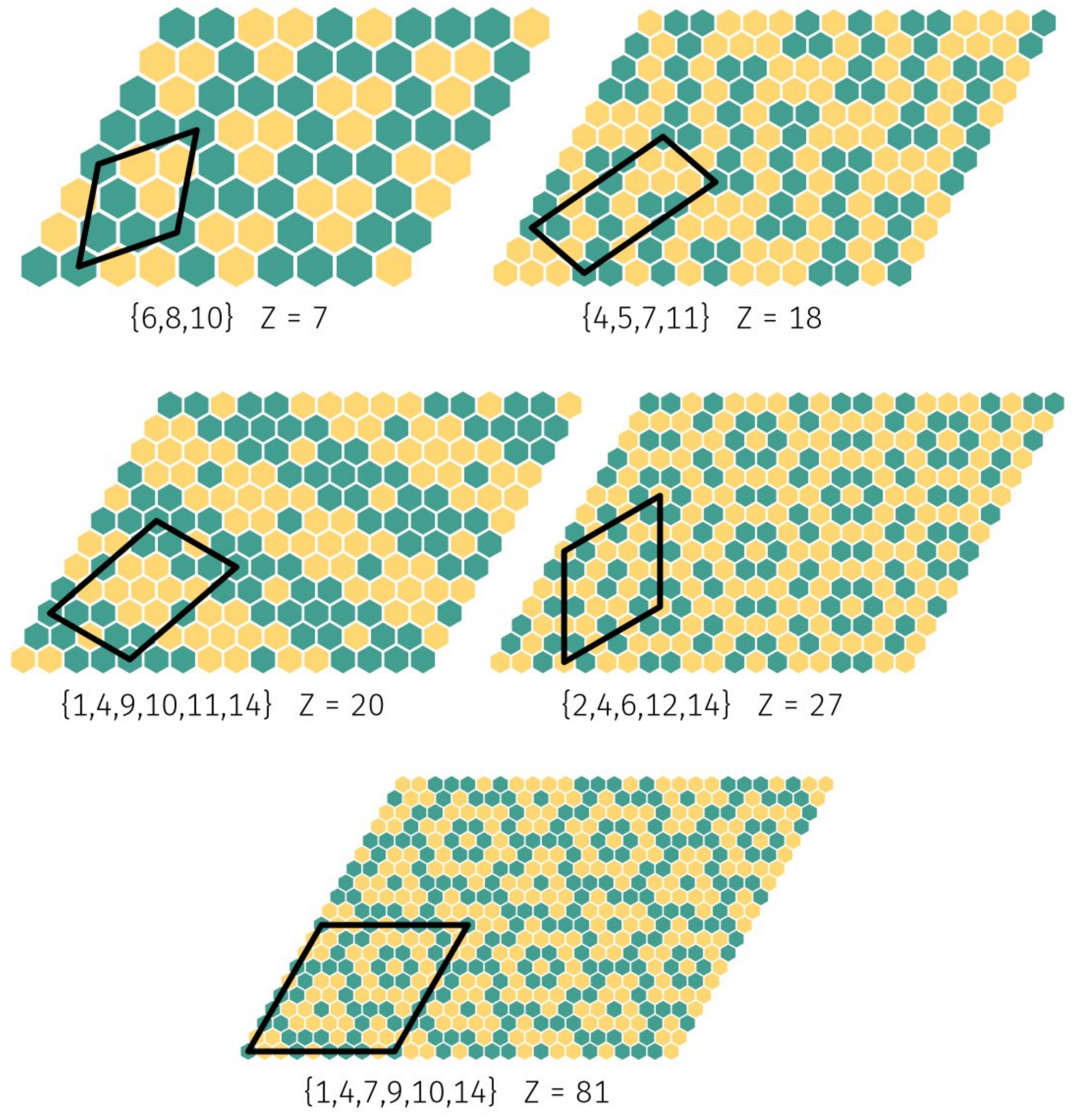

**Figure 3.** Examples of the crystalline structures with the unit cell indicated. Each structure is labelled by the list of constituent local structures (see text) and the cell size Z.

**3. The Diversity of Periodic and Non-Periodic Groundstates**

Our measure of structural diversity [10] is based on the Hill's numbers used in population ecology [24]. The idea is to characterise a configurations by the frequency of different local structures, our 'species'. In the case of the FLS model there are just the 14 local structures shown in Fig. 1. Let $p_i$ be the fraction of sites corresponding to local structure i. Then we can define the set of Hill's numbers $\{S_\alpha\}$ by

$$S_\alpha = \left( \sum_i^{S_0} p_i^\alpha \right)^{\frac{1}{1-\alpha}} \tag{2}$$

where $S_0$ is the total number of species present (i.e. $p_i > 0$) in a structure. The Hill's numbers are related to Renyi entropies [25] $H_\alpha$ via

$$H_\alpha = \ln S_\alpha \tag{3}$$

where $\alpha$ can take on an integer value 0, 1, 2, ··· ∞. With increasing value of the index $\alpha$, the associated diversity $S_\alpha$ is weighted more heavily towards the most frequent 'species'. If we set $\alpha = 0$, we have a diversity $S_0$ equal to the number of different species present, irrespective of the relative abundances. Such a measure is highly sensitive to the inclusion rare local structures. In the opposite limit, if we set $\alpha = \infty$, then we have a diversity $S_\infty = 1/p_{max}$, where $p_{max}$ is the frequency of the most abundant species. This represents a lower bound on diversity. Its chief virtue is that $p_{max}$ can be measured to reasonable accuracy using small samples. Previously [26], we have shown that an $\alpha > 0$ is useful to supress noise arising from rare states whose frequency fluctuates significantly with each generation of the groundstate. We shall use $S_1$ as our standard measure of diversity, given by Eq.3 with

$$H_1 = - \sum_i^{S_0} p_i \ln p_i \tag{4}$$

The quantity $H_1$ is analogous to the measure of structural information content, $I_G$, of a crystal proposed by Krivovichev [3] which is defined as follows. In a unit cell there are sets of positions that can be generated from a single member of the set through that application of the symmetry operations of the space group. The number of different sets is k and each set is identified by one representative member known as the Wyckoff position. The quantity $p_j$ is the fraction of sites in the unit cell that belong to the set represented by the Wyckoff position j. The structural information content $I_G$ [3] is then expressed as

$$I_G = -\sum_i^k p_i \ln_2 p_i \tag{5}$$

While $H_1$ and $I_G$ are similar in form, they are distinct. The set of Wyckoff positions used in Eq. 5 are determined by the specific symmetry elements of the crystal space group, while the set of local structures used in the structural diversity $S_1$ are determined by local topology alone.

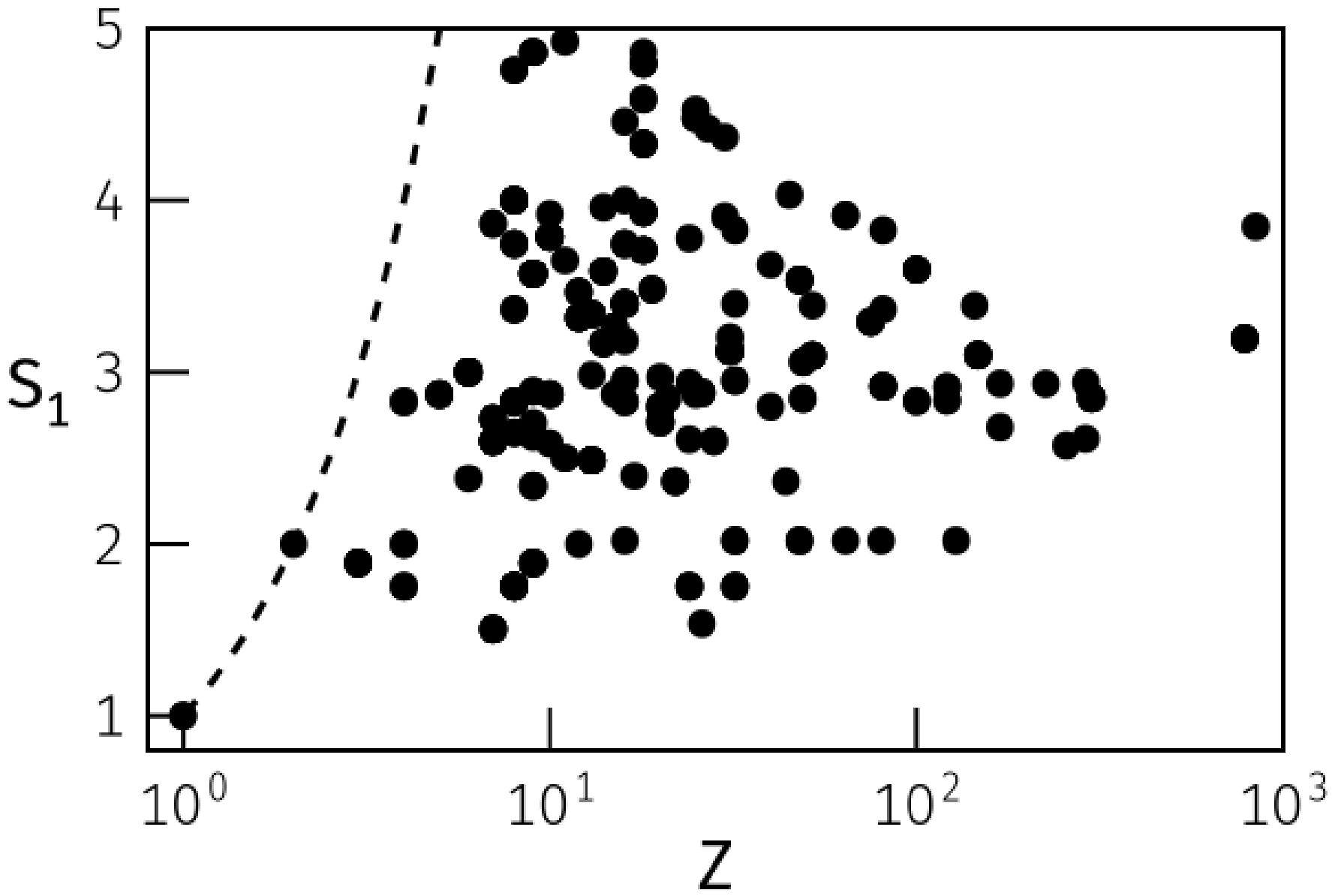

**Figure 4.** Scatter plot of the diversity $S_1$ as a function of the unit cell size logZ for the crystal groundstates of the FLS model. The dashed line represents the lower bound on Z, i.e. $Z \geq S_1$.

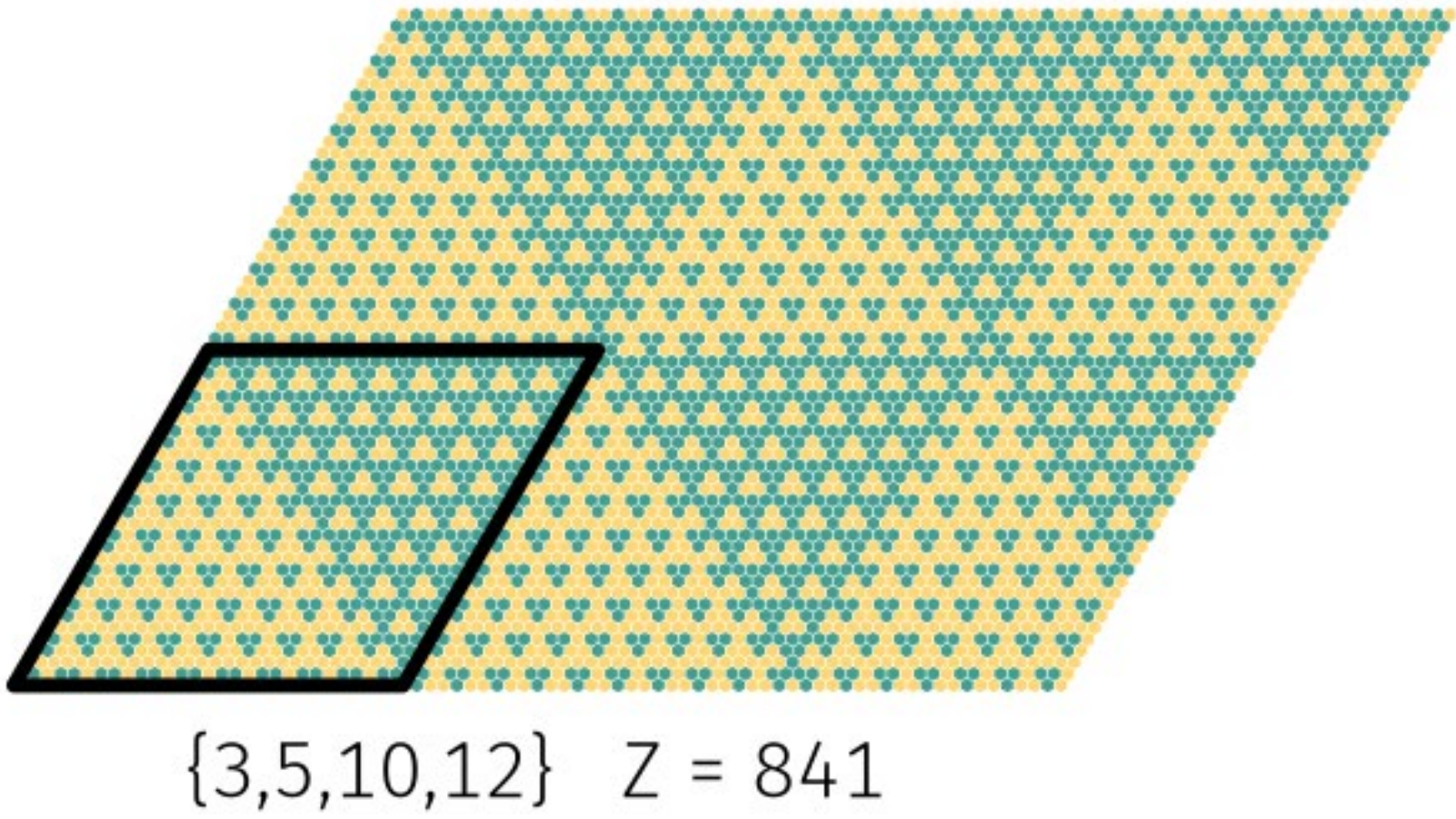

**Figure 5.** The periodic groundstate of a 'giant' unit cell structure. The list of constituent local structures and the unit cell size Z are indicated. The large magnitude of Z is a consequence of the combination of a simple crystal and its compositionally-inverted twin.

To start we can look at how diversity compares with the unit cell size Z of the periodic groundstates. In Fig. 4 we plot of the diversity $S_1$ against Z for the 287 periodic groundstate of the FLS model. We note that there is little correlation between $S_1$ and Z. The diversity does set a lower bound on the unit cell size, as shown, but that is all. This difference is not surprising. Where $S_1$ counts the number of distinct structural units in a unit cell, Z includes the number of repetitions of each unit, often in different orientations, required to assemble the repeating cell (see Fig. 5). We note that our 'giant' unit cells – i.e. those with $Z > 100$ – are dominated by diversities in the range $2 < S_1 < 3$, i.e. they are constructed from a relatively small set of local structures.

Turning to the complete set of groundstate structures, we plot, in Fig. 6, the distributions of diversity for the periodic and non-periodic groundstates. Two features of this plot are worth noting. The first is that the distribution of periodic groundstates is truncated at $S_1 \sim 5$. There is, it appears, an upper bound on the diversity of crystals. (A similar observation based on a smaller structural data set was reported in ref. [26]). This result for the FLS model is qualitatively consistent with the analysis of the experimental data for intermetallics. Daams and Villars [27] reported that 82% of all intermetallic crystals in the Inorganic Chemistry Data Base [28] had a diversity of 4 or less. We are not suggesting that the observed diversity threshold for crystallinity in a 2D lattice model is being replicated in the empirical data for the intermetallics. What the two observations do suggest is that there is a general reduction in crystalline stability with increasing diversity, a trend for which there is not, to our knowledge, any theoretical rationalization. Periodic order arises as a consequence of maximising the density of a small number of favored local structures. The results shown in Fig. 6 place a quantitative bound on what ‘a small number’ means in this statement. When the number of favored local structures exceeds this threshold, it appears that periodicity is no longer needed to construct the global minimum configurational energy.

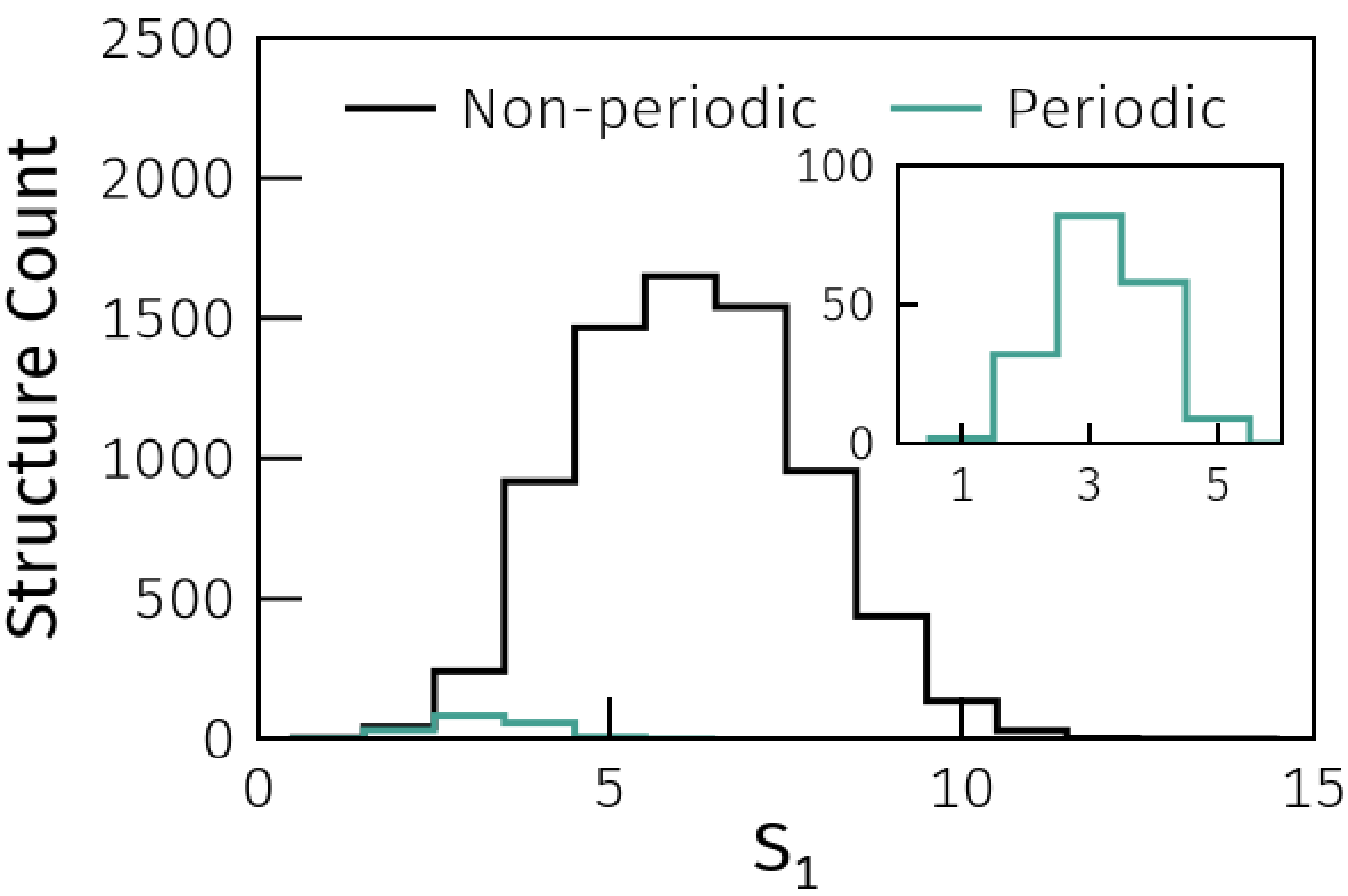

**Figure 6.** The distributions of $S_1$ for the periodic and non-periodic groundstates of the FLS model. An enlarged plot of the distribution of periodic structures is shown in the insert.

The second observation we can take away from Fig. 6 is the overwhelming predominance of non-periodic structures with diversities that span the complete range of possible values. Non-periodic structures make up 96% of the structures we can generate. Particularly striking, even when we restrict the diversity to $S_1 \leq 5$ (the range over which periodic structures occur), non-periodic structures still account for roughly 35% of the groundstates.

**4. Structural Selectivity and Order**

This huge host of non-periodic structures brings us to a core question of this paper - are there forms of order present in these non-periodic structures? To address this possibility, we need to identify a new signature of order, one that does not rely on recognizing some specific motif or regularity. As introduced in Section 1, the existence of an ordering principle implies that there are a set of structures that are incompatible with that principle and, hence, are excluded when the groundstate is formed, even when the local structure is stabilized by the Hamiltonian. This attribute of order is associated with the minimum free energy cost for introducing a new local structure into a given groundstate.

We shall identify whether a structure is selective or not via the following protocol. First, note that the space of systems (i.e. choices of FLS) is larger than the space of structures – 16384 systems as compared to the 7609 distinct structures we have found in the FLS model. This means that a structure may be the groundstate of a number of different systems. 'Selectivity' is an attribute we assign to groundstate structure, not to a system of chosen FLS's (i.e. the Hamiltonian). Each structure, identified by its specific set of local structures, can occur as the groundstate of more than one system (i.e. choice of FLS). If the structure occurs in at least one system where one or more FLS have been excluded from the structure, we designate the structure as selective. We note that the set of constituent local structures does not always uniquely specify a structure. It is possible for two distinct structures to share the same constituent local structures but in different proportions and different spatial arrangements. We believe that the proposed level of distinction between structures is sufficient for our purpose in this paper.

We shall start our analysis of selectivity with the periodic structures. In Fig. 7 we plot the distribution of diversity for selective and nonselective *periodic* structures. As expected, the majority of periodic structures are selective. What is interesting is the existence of a fraction of periodic structures that are not selective. How can an arbitrary local structure injected into a periodic structure not give rise to a high energy defect that the system would exclude on minimization? The answer lies in the proximity in the configurations space of the growing number of non-periodic structures to the periodic ones with increasing diversity. We find that the non-selective periodic structures generally form a non-periodic groundstate with the addition of the new FLS. Since a periodic structure is obviously ordered, the existence of non-selective periodic structures demonstrates that, while the presence of selectivity implies an underlying ordering principle, its absence does not exclude the possibility of order.

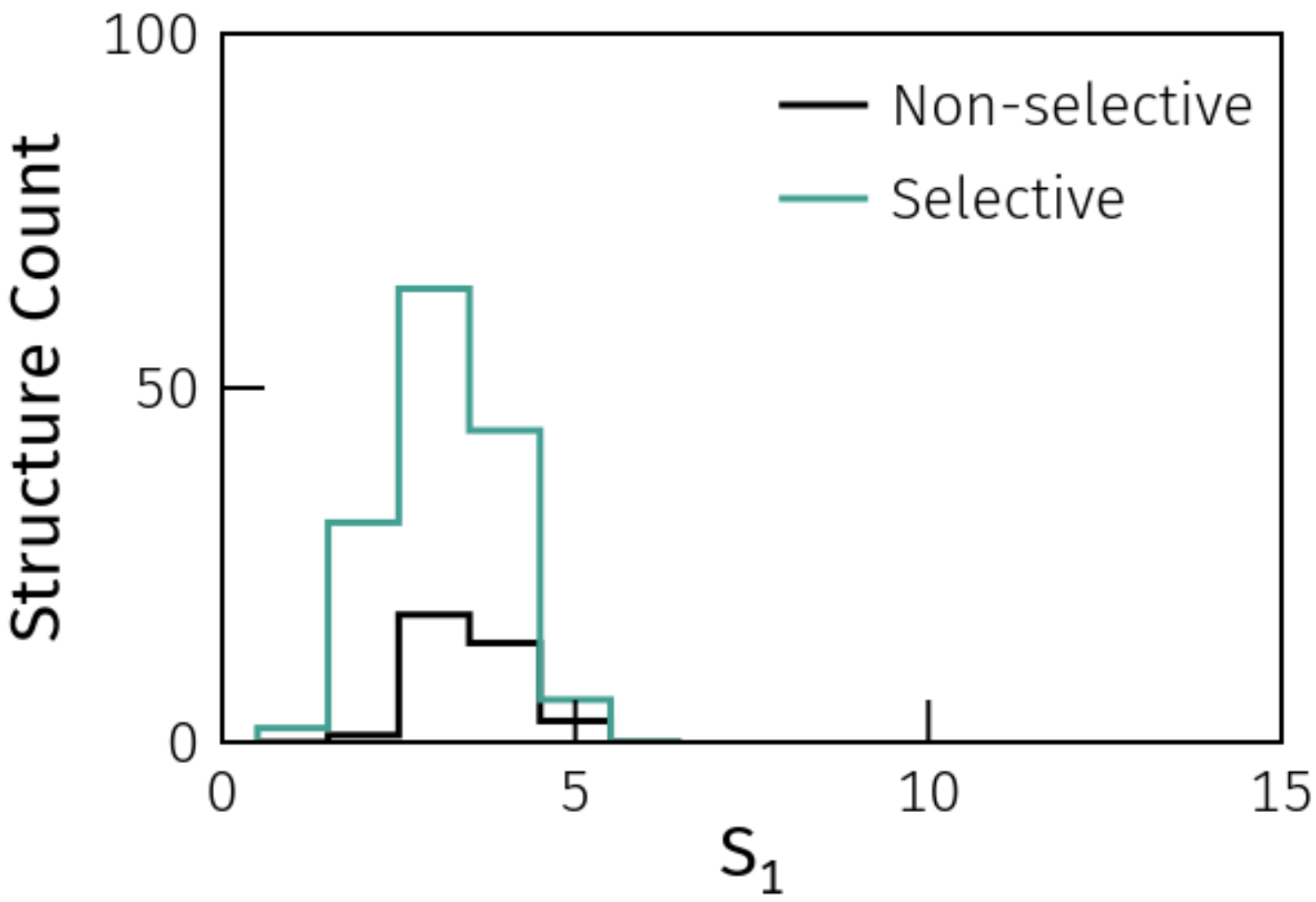

**Figure 7.** The distribution of diversity among selective and non-selective periodic groundstates of the FLS model.

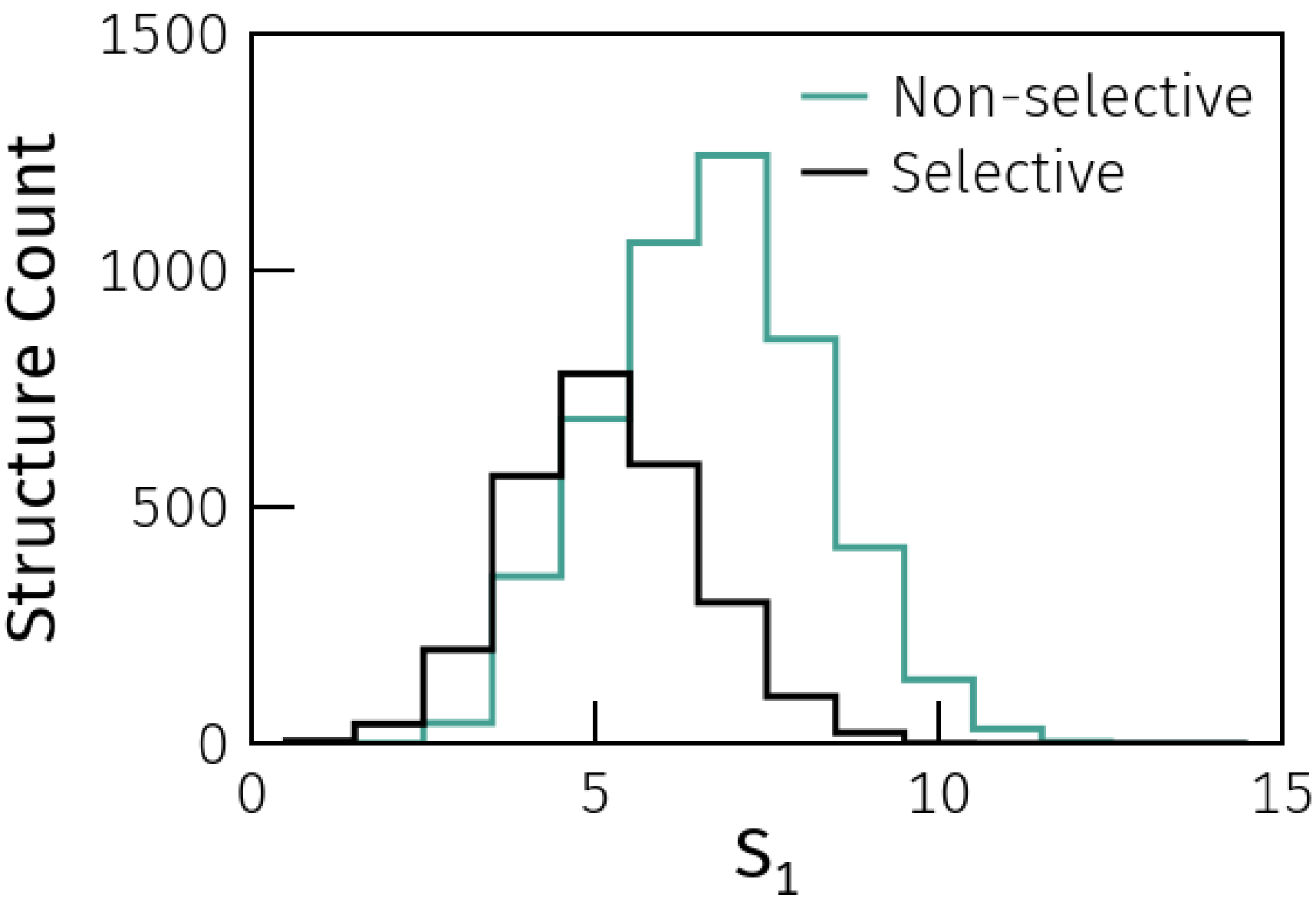

**Figure 8.** The distribution of $S_1$ for selective and non-selective non-periodic structures.

Now we come to address our central question – can we establish the existence of order in the population of non-periodic structures? In Fig. 8 we plot the distribution of diversity for the selective and nonselective non-periodic structures. What is striking is the substantial population of selective structures we find across a broad range of diversity. That selectivity can persist up to a diversity of $S_1 \sim 9$ is remarkable (remembering the periodic order disappears for $S_1 > 5$). The number of these selective non-periodic structures is also noteworthy. We have identified 2602 structures or 35% of all of the non-periodic groundstates as being selective. This is a major result of this paper – the demonstration that the absence of crystalline order does not exhaust the possibilities of order in condensed matter.

What types of structure are included in this category: non-periodic selective? There exists a substantial literature on random and aperiodic tilings [29,30] that would belong to this category. A random tiling consists of tilings, such as the square-triangle tiling [31,32], that can generate both non-periodic and periodic arrangements. The term 'aperiodic' tilings is generally reserved for those tiles that cannot form a periodic structure. Examples of aperiodic tilings include the tilings of Penrose [33] and the 'einstein' tile of Smith et al [34]. It is not clear whether random tilings cover the variety of structures observed in our non-periodic/selective class. For example, many random tilings exhibit a broken symmetry in the form of a long range orientational order (with some exceptions [35]) that is not obvious in all of our selective structures. We present a preliminary attempt at categorizing some structure types among the non-periodic selective class by identifying the following 4 categories of structures.

i) *Crystals with random pointwise inclusions*. As shown in Fig. 9a, these structures retain a clearly identified crystal structure but include random pointwise insertions. The stabilization of these structures is a consequence of adding the FLS that stabilize the specific 'defect'.

ii) *Crystals with random grain boundaries*. As for the previous category, we can see evidence of a crystalline lattice (see Fig. 9b) but now the FLS stabilizes one or more grain boundaries which are then able to proliferate through the structure.

iii) *Irregular assemblies of a motif*. This category one or more local motifs are repeated throughout the structures (see Fig. 9c) but without any obvious organization of these motifs over an intermediate length scale.

iv) *Random networks*. As shown in Fig. 9d, these structures comprise of a repeating local motif that can link up to form extended network or labyrinth patterns.

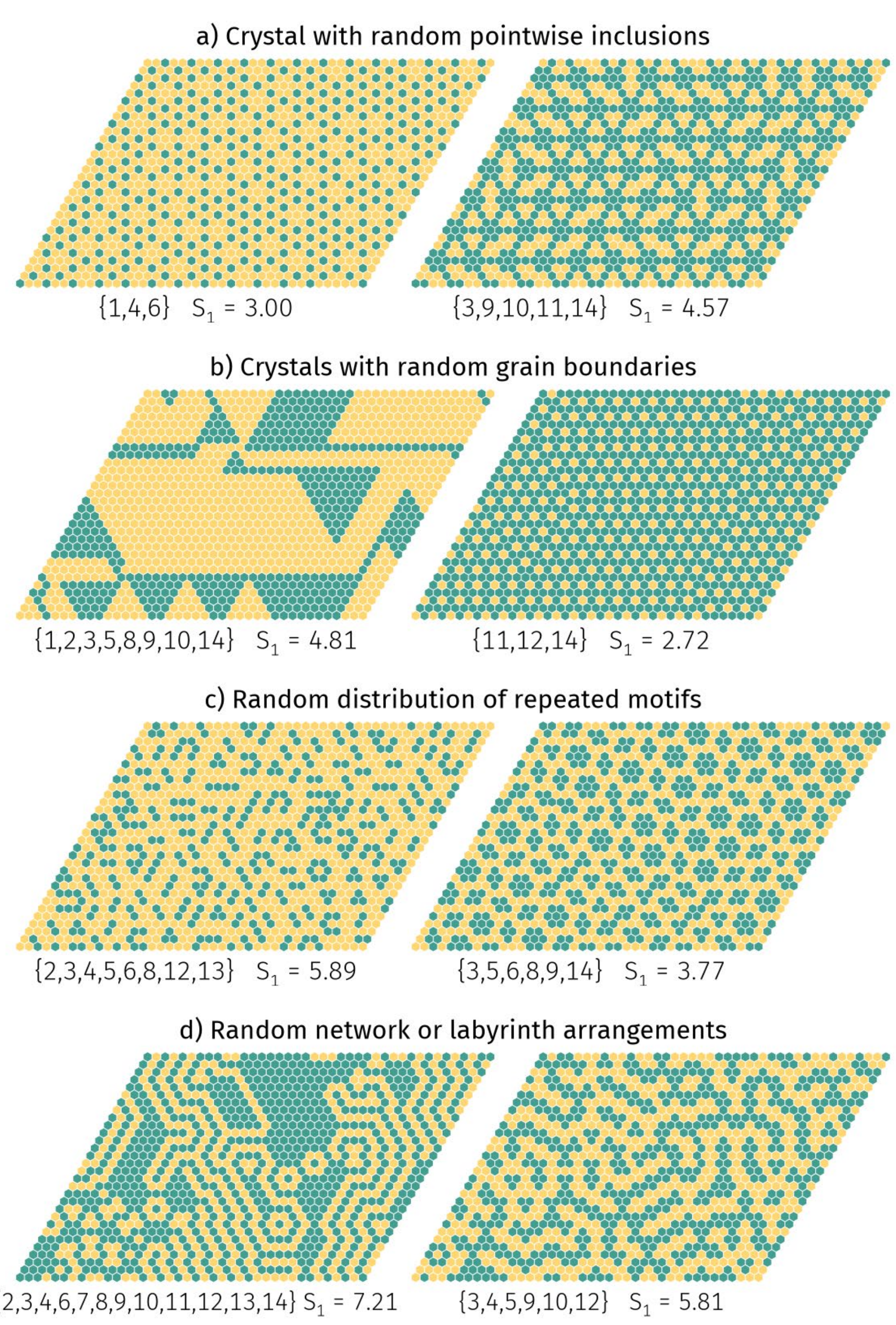

**Figure 9.** Examples of the four different types of non-periodic structures that exhibit selectivity as described in the text. The constituent local structures and the diversity $S_1$ for each structure is indicated.

We are left with 4824 structures that are neither periodic nor selective that we must, at least for now, lump together as ‘disordered’. We do so acknowledging that this designation is simply an admission that we have exhausted our current criteria for order. A significant fraction of these ‘disordered; structures have diversities $\leq 5$ , the range shared with the crystalline structures. In Fig. 10 we present a selection of these low diversity structures that we identify as non-periodic and non-selective. Their low diversity is immediately evident in each structure in form of repeated patterns and motifs. Despite these observable signs of organization, none of these structures exhibited any capacity to exclude an added favored local structures (hence their non-selective status) and so we cannot demonstrate the presence of an underlying ordering principle

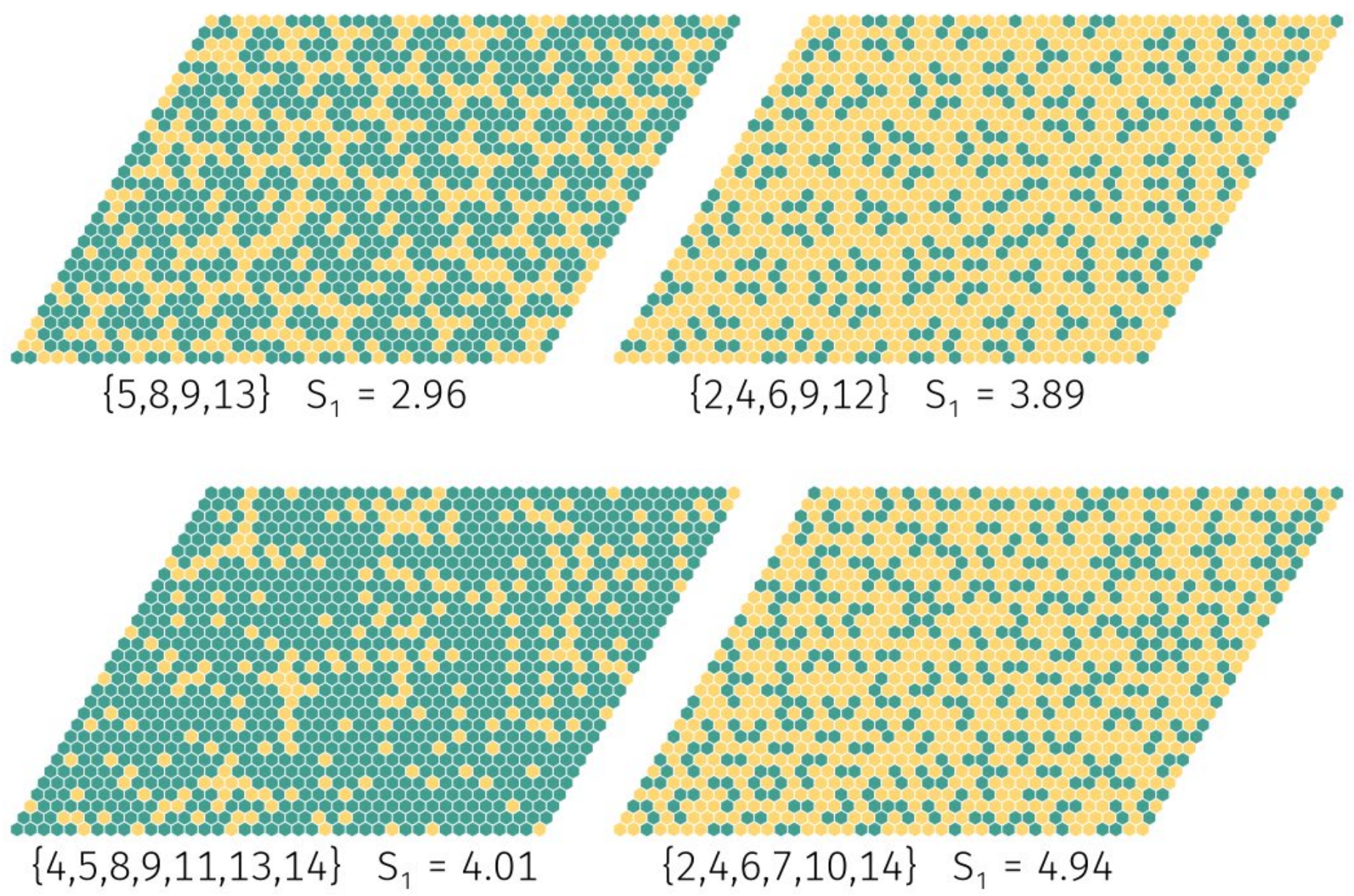

**Figure 10.** Four examples of low diversity structures identified as non-periodic and non-selective. The constituent local structures and the diversity $S_1$ for each structure is indicated.

**5. Structural Selectivity in Off-Lattice Models of Liquids**

It is of obvious interest to extend these studies to off-lattice models of liquids. Calculating the structural diversity of a liquid of interacting atoms presents no significant problem. We have previously reported just such calculations for binary alloys [10] using a Voronoi analysis to classify local coordination structures. Extending the concept of structural selectivity to more realistic systems, however, is not so straightforward. We propose the following methodology. Consider a system characterised by an interaction potential energy $V_o$. We can generate the groundstate and determine the local structures present via a quench protocol similar to that used in this paper. We can then identify the set of local structures that are present in the high temperature liquid but absent in the groundstate. These excluded structures are the ones we shall use to determine the selectivity of the groundstate associated with the original system (i.e. with the potential energy $V_0$). We select an excluded structure $\alpha$ from this set and then recalculate the groundstate but now for the modified potential $V_0+V_\alpha$, where $V_\alpha$ is a potential that specifically stabilizes the test structure $\alpha$. The use of modified potential energies in condensed matter simulations is already well established. The metadynamics algorithm, introduced by Laio and Parrinello [36], uses modified potential energies to steer configurations away from those already sampled. What we are suggesting is to use similar modifications to explore the susceptibility of groundstates to the inclusion of a nominated local structure. If the new groundstate is the same as the old, then we conclude that the groundstate is selective. If not, then we move on to the next excluded structure, construct a new $V_\alpha$. and repeat the quench. We plan to implement this approach in future work.

## 6. Conclusions

To talk about what lies between these two structural limits of crystalline and the completely disordered, we need a metric that allows us to define what we mean by 'between'. In this paper we have demonstrated that structural diversity provides this metric. In our calculation of the distribution of diversity over the set of 7609 groundstate structures generated by the FLS model, we have shown that the non-periodic groundstates predominate over the entire range of possible diversities. Whether this result is retained when we look at more realistic atomic or molecular interactions represents an important open question.

We have introduced a new signature of the presence of an ordering principle inherent in a given groundstate structure, structural selectivity. This property reflects the susceptibility of a structure to the inclusion of new local structures. We note that the resistance of a system to the inclusion of a new structure can arise from both energetic considerations (e.g. the new structure requires the inclusion of high energy structures) and entropic ones (e.g. the new structure can only accommodate a limited number of neighboring structures). As realized in this paper, measuring structural selectivity involves the manipulation of the Hamiltonian – a potentially awkward procedure which may not translate easily to all model systems. In the 2D lattice model, we have shown that roughly 35% of the large population of non-periodic groundstates exhibit selectivity and, hence, some form of underlying order. This is the major result of the paper. Unlike periodic order which, in the FLS model, is restricted to low diversities, $S_1 \leq 5$, the non-periodic order extends over a broad diversity range, $2 \leq S_1 \leq 9$. The identification of these non-periodic ordered structures is an important first step in addressing a number of very general questions about complex structures. Our preliminary suggestions regarding possible distinct types of non-periodic organization underscore the variety of ordering possibilities that might exist. The influence of diversity and selectivity on the thermodynamics of formation of these groundstates on cooling is a question we plan to address in future papers.

The term 'disorder', although widely used, provides us with no information about the structure of a state save that it does not conform to some specific set of target structures. This apophatic approach to order is perfectly adequate to describe the formation or disruption of a specific structure but it is of little use when we wish to compare the structures of two 'disordered' states. In this paper we have presented two quantities – diversity and selectivity – with the intent that they can usefully displace the term 'disorder' from the discussion of non-periodic structures. We hope that these attributes will contribute to the development of a more nuanced conceptual framework within which to analyse and understand complex material structures.

**References**

1. L. Pauling, The principles determining the structure of complex ionic crystals, J. Am. Chem. Soc. 51, 1010 (1929)

2. W. H. Baur, E. T. Tillmanns and W. Hofmeister, Topological analysis of crystal structures. Acta Cryst., B39, 669-674 (1983).

3. S. V. Krivovichev, Topological complexity of crystal structures: quantitative approach. Acta. Cryst. A68, 393-398 (2012).

4. T. Matsumoto and H. Wondratschek, Possible super-lattices of extraordinary orbits in 3-dimensional space, Zeitschrift Krist. 150, 181-198 (1979).

5. S. V. Krivovichev, Structural complexity of minerals: information storage and processing in the material world, Mineral. Mag, 77, 275-326 (2013).

6. J. Kurchan and D. Levine, Order in glassy systems, J. Phys A 44, 035001 (2011).

7. S. Martiniani, P. M. Chaikin and D. Levine, Quantifying hidden order out of equilibrium, Phys. Rev. X 9, 011031 (2019).

8. I. M. Douglass, J. C. Dyre and L. Costigliola, Complexity scaling of liquid dynamics, Phys. Rev. Lett. 133, 068001 (2024).

9. I. Fraenkel, J. Kurchan and D. Levine, Information and configurational entropy in glassy systems, arXiv:2402.05081v1 [cond-mat.dis-nn] (2024).

10. D. Wei, J. Yang, M. Q. Jiang, L. H. Dai, Y. J. Wang, J. C. Dyre, I. Douglass, and P. Harrowell, Assessing the utility of structure in amorphous materials, J. Chem. Phys. **150**, 114502 (2019).

11. L. Pi, F. Casanoves, and J. Di Rienzo, *Quantifying Functional Biodiversity* (Springer, Amsterdam, 2012).

12. Y. Q. Cheng and E. Ma, Atomic-level structure and structure-property relationship in metallic glass. Prog. Mater. Sci. 56, 379-473 (2011)

13. C. P. Royall and S. R. Williams, The role of local structure in dynamic arrest. Phys. Rep. 560, 1-75 (2015)

14. H. W. Sheng, W. K. Luo, F. M. Alamgir, J. M. Bai and E. Ma, Atomic packing and short-to-medium-range order in metallic glass. Nature 439, 419-425 (2006).

15. A. Malins, S. R. Williams, J. Eggers and C. P. Royall, Identification of structure in condensed matter with topological cluster classification. J. Chem. Phys. 139, 234506 (2013).

16. M. Leocmach and H. Tanaka, Role of icosahedral and crystal-like order in the hard spheres glass transition. Nat. Comm. 3, 974 (2012).

17. R. Lorand, *Aesthetic Order: The Philosophy of Order, Beauty and Art* (Routledge, London, 2000)

18. W. Steurer, Quasicrystals: What do we know? What do we want to know? What can we know? Acta Cryst A74, 1-11 (2017).

19. F. Gähler, Matching rules for quasicrystals: the composition-decomposition methods, J. Non-Cryst. Solids 153-154, 160-164 (1993).

20.W. Li, H. Park and M. Widom, Phase diagram of a random tiling quasicrystal, J. Stat. Phys. 66, 1- 68 (1992)

21. F. Gähler, Cluster coverings as an ordering principle for quasicrystals, Mater. Sci Eng. 294-296, 199-204 (2000).

22. P. Ronceray and P. Harrowell, The variety of ordering transitions in liquids characterized by a locally favoured structure, Europhys. Lett. **96**, 36005 (2011).

23. P. Ronceray and P. Harrowell, Favoured local structures in liquids and solids: a 3D lattice model, Soft Matter **11**, 3322 (2015).

24. M. O. Hill, Diversity and evenness: a unifying notation and its consequences. Ecology 54, 427-432 (1973).

25. A. Renyi, On measures of entropy and information. Proc. of the 4th Berkeley Symposium on Mathematical Statistics and Probability, ed. J. Neyman (University of California Press, 1961), pp. 547–561.

26. Y. Wang and P. Harrowell, Structural diversity in condensed matter: a general characterization of crystals, amorphous solids and the structures between. J. Chem. Phys. 161, 074502 (2024).

27. J. L. C. Daams and P. Villars, Atomic environment classification of the tetragonal ‘intermetallic’ structure types, J. Alloys Comp. 252, 110-142 (1997).

28. G. Bergerhoff, R. Hundt, R. Sievers and I. D.Brown, The inorganic crystal structure data base. J. Chem. Inf. Comput. Sci. **23** (2): 66–6910 (1983).

29. B. Grünbaum and G. C. Shephard, *Tilings and patterns* (W.H. Freeman, New York, 1987).

30. C. Richard, M. Hoeffe, J. Hermisson and M. Baake, Random Tilings: Concepts and Examples, J. Phys. A: Math. Gen. 31, 6385-6408 (1998).

31. A. Widmer-Cooper and P. Harrowell, Structural phases in non-additive soft-disk mixtures: Glasses, substitutional order, and random tilings, J. Chem. Phys. 135, 224515 (2011).

32. E. Tondl, M. Ramsay, P. Harrowell and W. Widmer-Cooper, Defect-mediated relaxation in the random tiling phase of a binary mixture: Birth, death and mobility of an atomic zipper, J. Chem. Phys. 104, 104503 (2014).

33. R. Penrose, Remarks on Tiling: details of a $1 + \varepsilon + \varepsilon^2$-aperiodic set. *The Mathematics Long Range Aperiodic Order*, NATO Adv. Sci. Inst. Ser. C. Math. Phys. Sci. **489**: 467–497 (1997).

34. D. Smith, J. S. Myers, S. Craig and C. Goodman-Strauss, An aperiodic monotile. Comb. Theory 4(1). http://dx.doi.org/10.5070/C64163843 (2024).

35. D. Charrier and F. Alet, Phase diagram of an extended classical dimer model, Phys. Rev. B 82, 014429 (2010).

36. A. Laio and M. Parrinello, Escape from free energy minima, Proc. Nat. Acad. Sci. USA 99, 12562-12566 (2002).